\pdfoutput=1
\documentclass{PoS}
\usepackage[lofdepth, lotdepth]{subfig}
\usepackage{caption}
\usepackage{microtype}
\usepackage[utf8]{inputenc}
\usepackage[T1]{fontenc}
\usepackage[version=3]{mhchem}
\usepackage[capitalize]{cleveref}

\newcommand*{\mum}{\ce{$\mu$-}}

\newcommand*{\numu}{\ce{$\nu$_{$\mu$}}}

\newcommand*{\cczpi}{CC0\ce{$\pi$}}
\newcommand*{\ccopi}{CC1\ce{$\pi$+}}

\newcommand*{\pd}{P\O{}D}

\title{A measurement of the \numu{} charged-current cross section
  on water with zero pions in the final state at T2K}

\ShortTitle{Muon neutrino \cczpi{} cross section}

\author{\speaker{Tianlu Yuan}\\
        University of Colorado Boulder\\
        E-mail: \email{tianlu.yuan@colorado.edu}}

      \abstract{The Tokai to Kamioka (T2K) experiment is a 295-km
        long-baseline neutrino experiment aimed towards the
        measurement of neutrino oscillation parameters $\theta_{13}$
        and $\theta_{23}$. Precise measurement of these parameters
        requires accurate knowledge of neutrino cross sections. We
        present a flux-averaged double differential measurement of the
        charged-current cross section on water with zero pions in the
        final state using the T2K off-axis near detector, ND280. A
        selection of $\nu_\mu$ charged-current events occurring in the
        Pi-Zero subdetector (\pd{}) of ND280 is performed with
        $5.8 \times 10^{20}$ protons on target. The charged, outgoing
        tracks are required to enter and be identified by the ND280
        Tracker. The cross section is determined using an unfolding
        technique. By separating the dataset into time periods when
        the \pd{} water layers are filled with water and when they are
        empty, a subtraction method provides a distribution of
        $\nu_\mu$ interactions on water only. Systematic uncertainties
        on the neutrino flux, interaction model, and detector
        simulation are propagated numerically within the unfolding
        framework.  }

\FullConference{38th International Conference on High Energy Physics\\
                  3-10 August 2016\\
                 Chicago, USA}

\begin{document}

\section{Neutrino interactions}
At the T2K peak energy of approximately 0.6 GeV, the primary
interaction mode is charged-current quasielastic (CCQE). In a CCQE
interaction on free nucleon targets, a neutrino (antineutrino)
interacts with neutron (proton) to produce its corresponding lepton
and a proton (neutron). With nuclear targets, complications arise from
both initial state nuclear effects and final state interactions
(FSIs). These intranuclear effects can alter the outgoing particle
kinematics and topologies. Pion absorption for example can make a
non-CCQE interaction appear as a CCQE final state. Our detectors are
not capable of detecting interactions within the nucleus and we are
forced to classify events based on the detectable, post-FSI
topology. A neutrino-nucleus interaction with a single muon and zero
pions is a CCQE-like topology called ``\cczpi{}''. Here, we present a
double differential measurement in $(p_\mu, \cos \theta_\mu)$ of the
\cczpi{} cross section on water using the \pd{} and tracker of ND280.

\section{Analysis strategy}
The \pd{} contains a water-target region of alternating layers of
water and hydrocarbon scintillator~\cite{t2k2012p0dnim}. The water
layers are passive but can be drained or filled during different
data-taking periods. \cczpi{} \numu{} interactions are selected for by
searching for the \mum{} signature. Only interactions consistent with
the neutrino beam timing are considered. We require the \mum{}
candidate to originate in the \pd{} water-target fiducial volume and
enter the tracker directly downstream of the
\pd{}~\cite{t2k2011nim}. The tracker contains TPCs which provide
accurate momentum reconstruction and allow us to select only
negatively charged tracks. To enhance the \cczpi{} signal, events with
multiple reconstructed objects in the \pd{} are excluded as they often
correspond to multi-pi interactions.
% event display

Two control samples (sidebands) are used to constrain the two largest
background topologies in the signal selection. These are events with
either a single pion, \ccopi{}, or any other charged-current neutrino
interaction, CCOther. Events with a \mum{} candidate and extra tracks
in the \pd{} are used for sideband selections. If, including the
\mum{} candidate, the \pd{} reconstructed exactly two tracks and a
Michel electron from stopped muon decay the event is classified as
\ccopi{}. If, including the \mum{} candidate, the \pd{} reconstructed
more than two tracks the event is classified as CCOther. The ratio of
the overall data sideband normalization to the overall MC sideband
normalization is calculated and used to constrain the corresponding
background in the \cczpi{} MC selection.

The event selections described above are binned in the reconstructed
double differential $(p_\mu, \cos \theta_\mu)$ phase space. Detector
reconstructed variables are imperfect approximations to the muon's
true initial state. To extract the true kinematics from the
reconstructed, an unfolding technique is
used~\cite{dagostini1995bayesunf}. The purpose of unfolding is to
remove detector related imperfections to achieve a more accurate
representation of how the muon emerged from the interaction.

To measure interactions on water, the \pd{} was designed to be drained
and filled during different run periods. All things being equal except
for the inclusion or exclusion of water, a subtraction of the true
water-in and water-out distributions should give the number of
interactions on water. The idea for a measurement on water is thus to
first unfold the reconstructed distribution for water-in and water-out
separately to get an approximation of their true distributions, then
subtract the post-unfolding results to get the distribution that
occurred on water. The number of interactions on water is given as,
\begin{equation}
  \label{eq:unfoldsubtract}
  N_{i}^{O} = \frac{U_{ij}^{w} N_{j}^{w}}{\epsilon_{i}^{w}} - R
  \frac{U_{ij}^{a} N_{j}^{a}}{\epsilon_{i}^{a}},
\end{equation}
where the indexes $i$ and $j$ indicate true and reconstructed bins
respectively, $w$ and $a$ indicate water-in and water-out periods
respectively, $N$ is the number of purity-corrected, signal events
measured in the data signal selection, $\epsilon$ the selection
efficiency, and $R$ the flux normalization factor between water-in and
water-out periods. $U_{ij}$ represents the unfolding matrix. From
this, the differential cross section on water can be expressed as,
\begin{equation}
  \label{eq:diffxsec}
  \frac{d\sigma}{dx} = \frac{N_{i}^{O}}{F^{w} N_{n} \Delta_{i}},
\end{equation}
where $F^{w}$ is the integrated flux over the water-in period, $N_{n}$
the number of neutrons, and $\Delta_{i}$ the area of bin $i$ across
variable $x$. $F^w$ is calculated using flux simulations from FLUKA and
constraints from the NA61/SHINE hadron production experiment at
CERN~\cite{t2k2013flux}. As the water-in and water-out periods have
different beam exposures, total flux for the water-out periods, $F^a$,
is used to scale the flux normalization ratio, $R = F^w/F^a$.

\section{Systematic uncertainties}
Sources of systematic uncertainties affecting this measurement include
uncertainties on the flux, interaction model, and detector
simulation. Flux uncertainties are due in large part to uncertainties
in the hadron production model but are affected by beamline
uncertainties as well. A parameterization in neutrino energy and
flavor is used to propagate flux uncertainties. Interaction model
uncertainties affect the interaction cross section and
FSI\@. Parameters that govern the cross section and FSI models are
used to propagate interaction uncertainties. Detector simulation
systematics affect the reconstructed kinematics of the \mum{}
candidate and selection efficiencies and purities. Uncertainties on
the \pd{} simulation include the fiducial volume water mass,
out-of-fiducial-volume contamination, and the \pd{} momentum
resolution. Uncertainties on the tracker simulation include magnetic
field distortions, TPC reconstruction efficiency, and the tracker
momentum resolution. Detector systematics are typically applied to the
MC reconstruction based on the MC truth information. All uncertainties
are propagated by reweighting or varying events in the nominal
simulation based on a perturbation of the underlying parameter. The
tweaked MC is then used to recalculate elements in
\cref{eq:unfoldsubtract,eq:diffxsec} which give tweaked cross section
results. Correlations between the water-in and water-out periods were
taken into account in the subtraction. A set of such results,
generated from a set of hypotheses on the MC, is used to construct the
error envelope on the final result.

% The subtraction of water-in and water-out samples can complicate the
% propagation of systematics onto the final result. Since flux and model
% uncertainties are independent of the detector configuration, they are
% treated as completely correlated between the water-in and water-out
% data-taking periods. Detector systematics are treated similarly,
% except that certain uncertainties, such as the water mass, only affect
% the water-in period. Statistical uncertainties are treated as
% uncorrelated and therefore make up a large fraction of the error on
% the subtracted result.

\section{Results}
\begin{figure}[!htb]
  \centering
  \includegraphics[width=\textwidth]{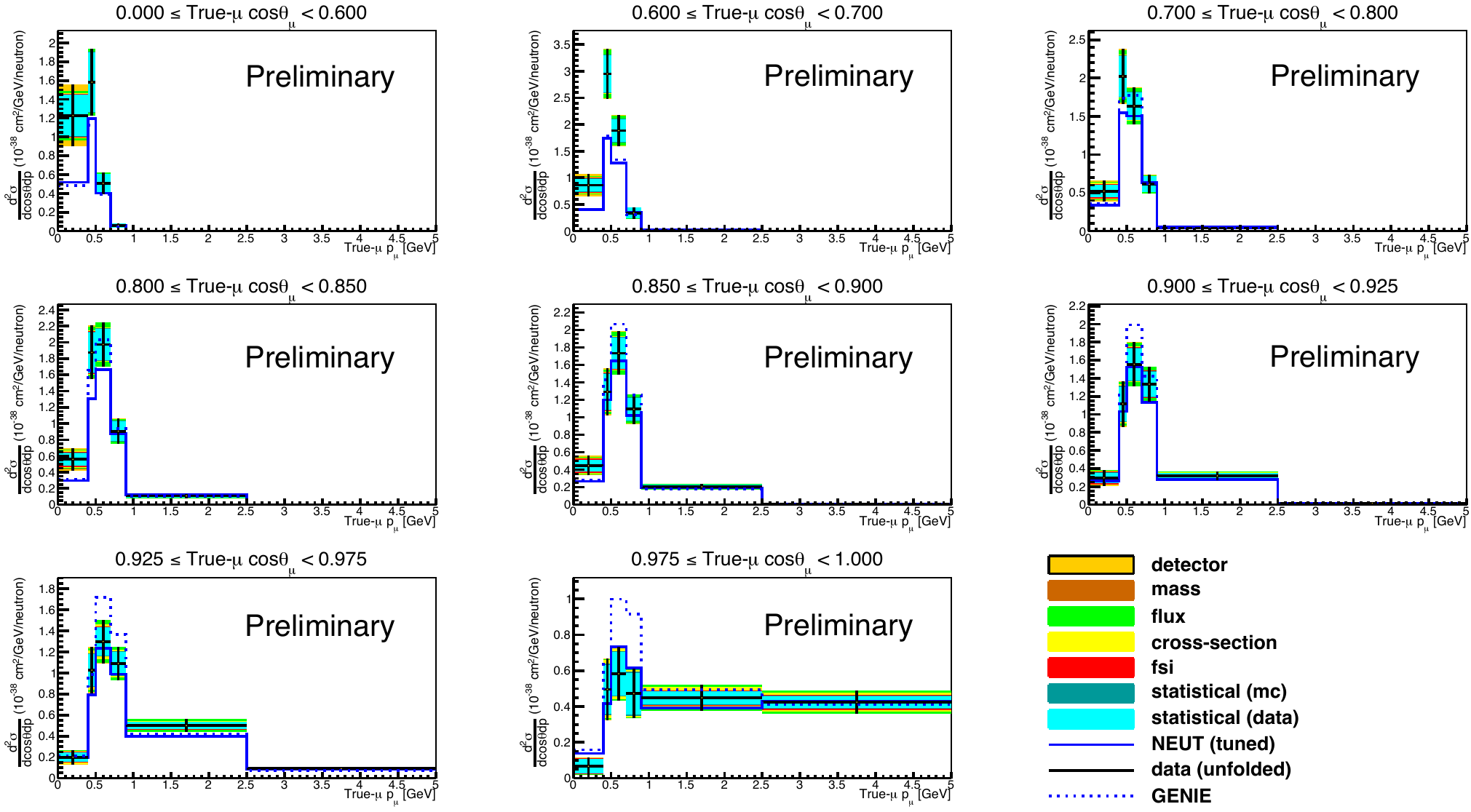}
  \caption[Result]{The double differential \cczpi{} water cross
    section.}
\label{fig:result}
\end{figure}

\begin{figure}[!htb]
  \centering
  \subfloat{
    \includegraphics[width=0.32\textwidth]{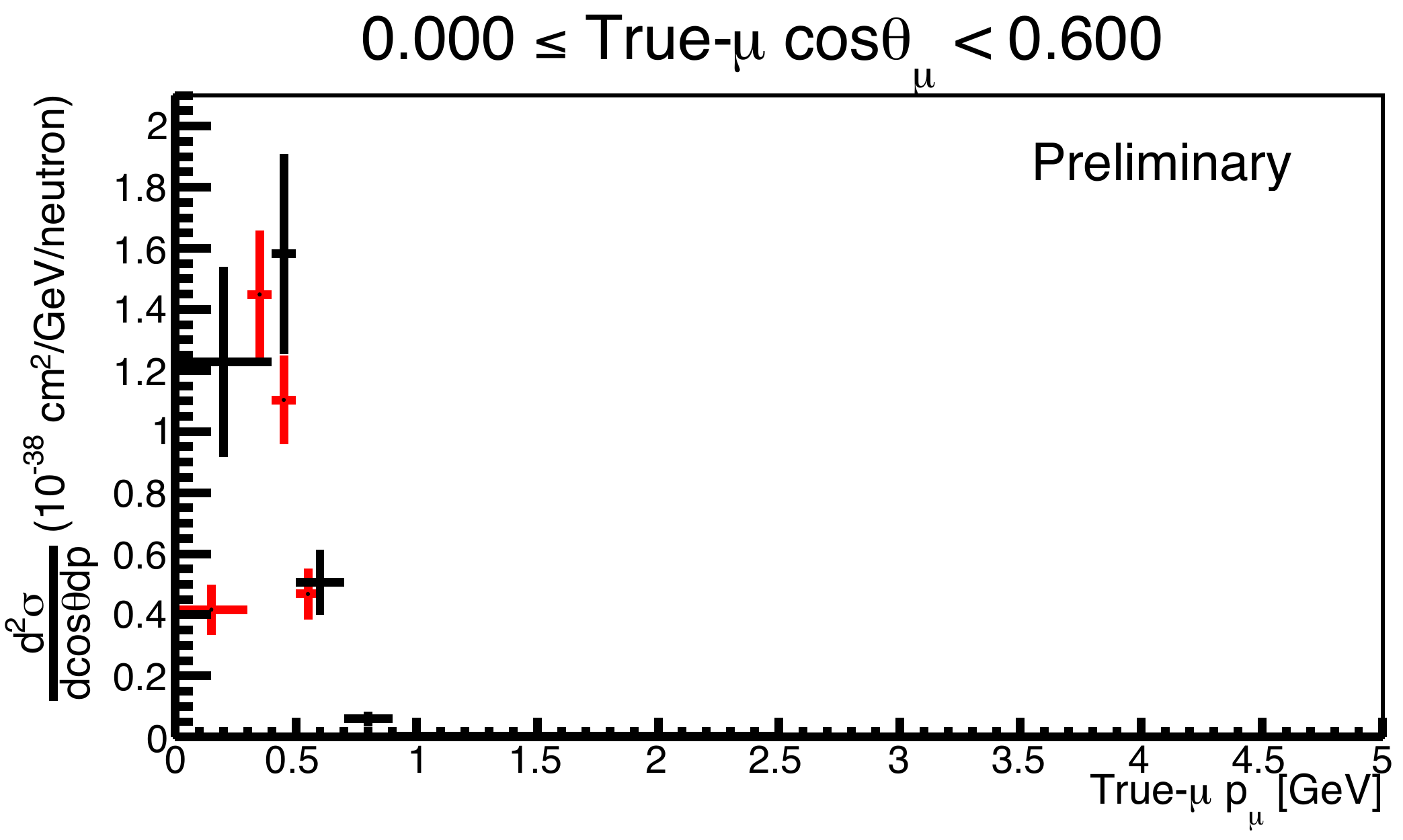}
  }
  \subfloat{
    \includegraphics[width=0.32\textwidth]{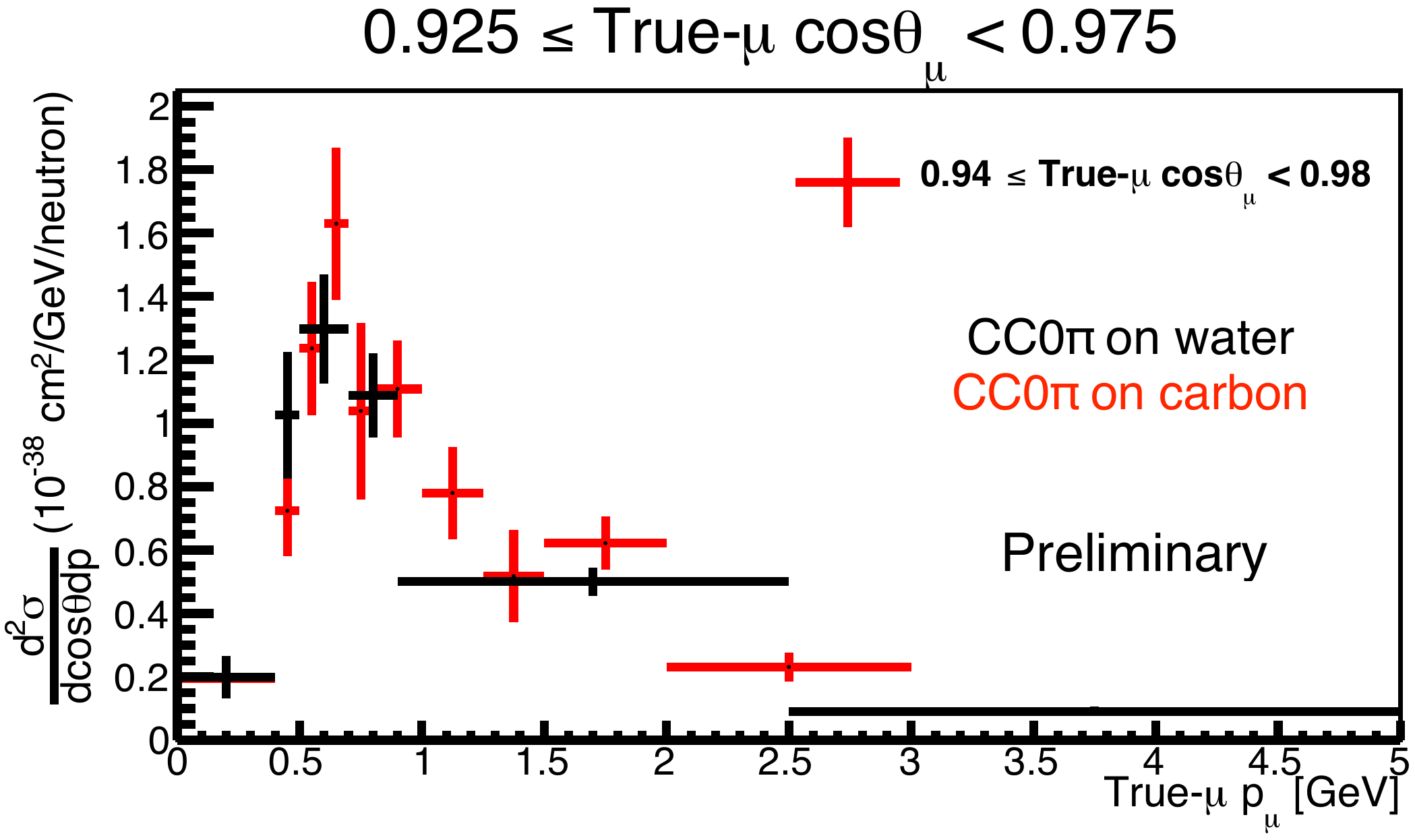}
  }
  \subfloat{
    \includegraphics[width=0.32\textwidth]{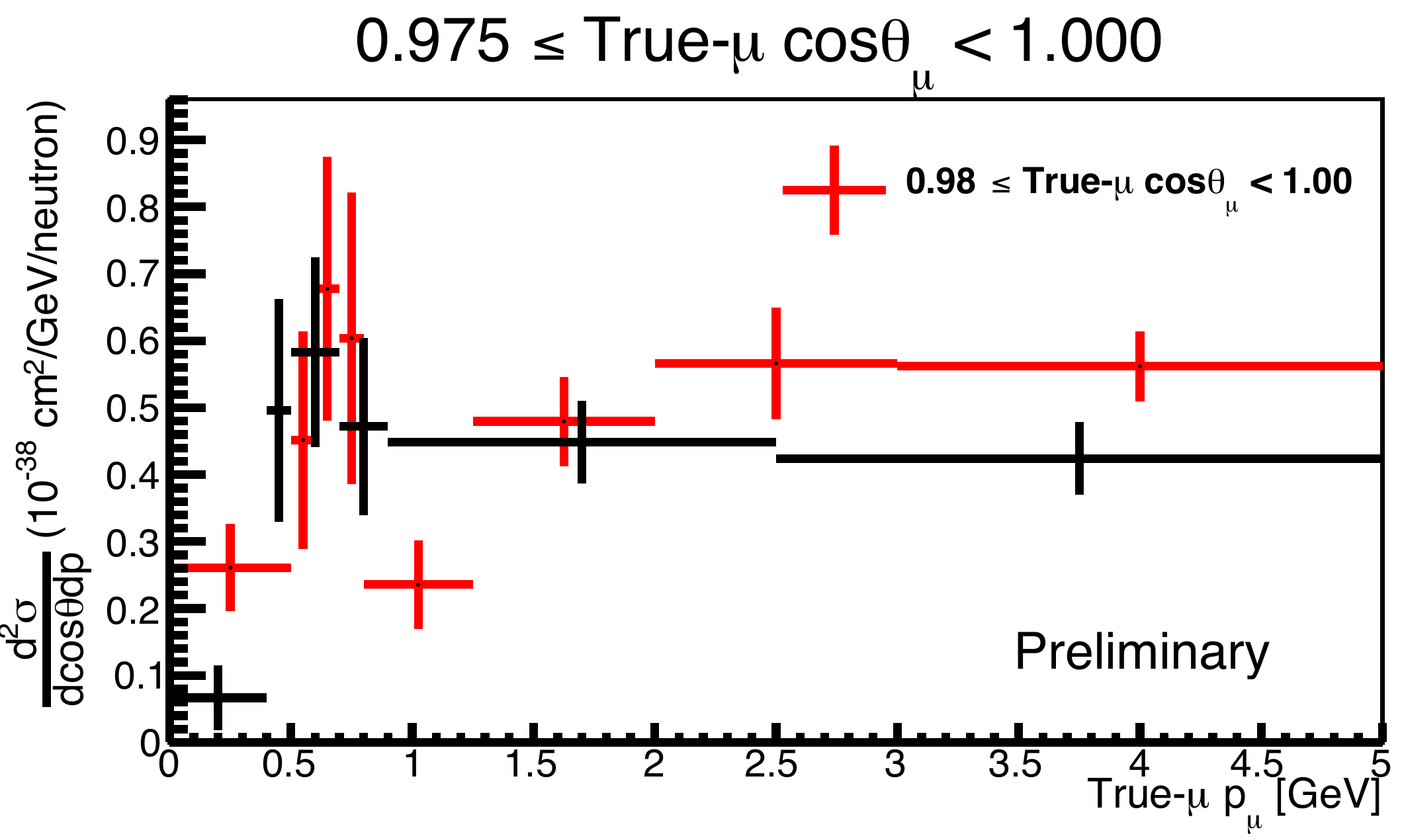}
  }
  \caption[Comparison to result on C]{A comparison of this result to
    that obtained in analysis 1 of \cite{t2k2016cczpicarbon} which
    reports a \cczpi{} measurement on carbon. Three angular slices are
    shown. Note that the binnings are not identical.}
\label{fig:compc}
\end{figure}

The results shown in \cref{fig:result} use data from T2K Runs
2--4. The colored bars show the cumulative error
contributions from various sources of uncertainty. Each error source
listed in the legend is treated independently and their uncertainties are
summed in quadrature. The black data points show the double
differential result with full errors. MC predictions from NEUT v5.3.2
tuned to a relativistic Fermi gas model with Random Phase
Approximation and the default GENIE v2.8.0 are shown as solid and
dashed blue lines. This result shows good agreement with the measured
\cczpi{} cross section on carbon from an earlier tracker-based T2K
analysis as shown in \cref{fig:compc}~\cite{t2k2016cczpicarbon}.

\begin{thebibliography}{99}
  \bibitem{t2k2012p0dnim}
    S. Assylbekov \emph{et.\ al.},
    \emph{The T2K ND280 off-axis pi–zero detector},
    \emph{Nucl. Instrum. Methods} {\bf A686} (2012) 48

  \bibitem{t2k2011nim}
    K. Abe \emph{et.\ al.},
    \emph{The T2K experiment},
    \emph{Nucl. Instrum. Methods} {\bf A659} (2011) 106

  \bibitem{dagostini1995bayesunf}
    G. D'Agostini,
    \emph{A multidimensional unfolding method based on Bayes' theorem},
    \emph{Nucl. Instrum. Methods} {\bf A362} (1995) 487

  \bibitem{t2k2013flux}
    K. Abe \emph{et.\ al.},
    \emph{T2K neutrino flux prediction},
    \emph{Phys. Rev.} {\bf D87} (2013) 012001
    {\tt arXiv:1602.03652 [hep-ex]}

  \bibitem{t2k2016cczpicarbon}
    K. Abe \emph{et.\ al.},
    \emph{Measurement of double-differential muon neutrino
      charged-current interactions on C$_8$H$_8$ without pions
      in the final state using the T2K off-axis beam},
    \emph{Phys. Rev.} {\bf D93} (2016) 112012
    {\tt arXiv:1602.03652 [hep-ex]}
\end{thebibliography}
\end{document}